\documentclass[]{spie}  
\usepackage[]{graphicx}
\usepackage[]{amsmath,amssymb,color}

\newcommand{\Z}{\mathbb{Z}}

\newcommand{\R}{\mathbb{R}}

\newcommand{\supp}{\text{supp}}

\def\minim{\mathop{\hbox{minimize}}}
\newtheorem{theorembf}[theorem]{\textbf{Theorem}}
\newtheorem{lemmabf}[theorem]{\textbf{Lemma}}
\newtheorem{propbf}[theorem]{\textbf{Proposition}}

\newtheorem{defn}{Definition}
\newtheorem{rem}{Remark}

\title{Weighted-{\LARGE$\ell_1$} minimization with multiple weighting sets} 

\author{Hassan Mansour\supit{a,b} and {\"O}zg{\"u}r Y{\i}lmaz\supit{a}
\skiplinehalf
\supit{a}Mathematics Department, University of British Columbia, Vancouver - BC, Canada; \\
\supit{b}Computer Science Department, University of British Columbia, Vancouver - BC, Canada
}

\authorinfo{Further author information: (Send correspondence to Hassan Mansour)\\Hassan Mansour: E-mail: hassanm@cs.ubc.ca;\\  {\"O}zg{\"u}r Y{\i}lmaz: E-mail: oyilmaz@math.ubc.ca}

\begin{document} 
  \maketitle 

\begin{abstract}
  In this paper, we study the support recovery conditions of weighted
  $\ell_1$ minimization for signal reconstruction from compressed
  sensing measurements when multiple support estimate sets with
  different accuracy are available. We identify a class of signals for
  which the recovered vector from $\ell_1$ minimization provides an
  accurate support estimate. We then derive stability and robustness
  guarantees for the weighted $\ell_1$ minimization problem with more
  than one support estimate. We show that applying a smaller weight to
  support estimate that enjoy higher accuracy improves the recovery
  conditions compared with the case of a single support estimate and
  the case with standard, i.e., non-weighted, $\ell_1$
  minimization. Our theoretical results are supported by numerical
  simulations on synthetic signals and real audio signals.
\end{abstract}




\section{INTRODUCTION}
\label{sec:intro}  
A wide range of signal processing applications rely on the ability to
realize a signal from linear and sometimes noisy measurements. These
applications include the acquisition and storage of audio, natural and
seismic images, and video, which all admit sparse or approximately
sparse representations in appropriate transform domains.

Compressed sensing has emerged as an effective paradigm for the
acquisition of sparse signals from significantly fewer linear
measurements than their ambient dimension \cite{Donoho2006_CS, CRT05,
  CRT06}. Consider an arbitrary signal $x \in \R^N$ and let $y \in
\R^n$ be a set of measurements given by
$$y = Ax + e,$$
where $A$ is a known $n\times N$ measurement matrix, and $e$ denotes
additive noise that satisfies $\|e\|_2\leq \epsilon$ for some known
$\epsilon\ge 0$. Compressed sensing theory states that it is possible
to recover $x$ from $y$ (given $A$) even when $n \ll N$, i.e., using
very few measurements.

When $x$ is strictly sparse, i.e. when there are only $k < n$ nonzero
entries in $x$, and when $e=0$, one may recover an estimate $x^*$ of
the signal $x$ as the solution of the constrained $\ell_0$
minimization problem
\begin{equation}\label{eq:L0_min}
\minim_{u\in \R^N}\ \|u\|_0 \ \text{subject to} \ \ Au=y.
\end{equation}
In fact, using \eqref{eq:L0_min}, the recovery is exact when $n \ge 2k$
and $A$ is in general position \cite{Donoho03}. However, $\ell_0$
minimization is a combinatorial problem and quickly becomes
intractable as the dimensions increase. Instead, the convex relaxation
\begin{equation}\label{eq:L1_min}
 \minim_{u \in \R^N}\ \|u\|_1 \ \text{subject to} \ \|Au - y\|_2 \leq \epsilon
\end{equation}
can be used to recover the estimate $x^*$.  Cand{\'e}s, Romberg and
Tao \cite{CRT05} and Donoho \cite{Donoho2006_CS} show that if $n
\gtrsim k\log(N/k)$, then $\ell_1$ minimization \eqref{eq:L1_min} can
stably and robustly recover $x$ from inaccurate and what appears to
be ``incomplete'' measurements $y = Ax + e$, where, as before, $A$ is
an appropriately chosen $n \times N$ measurement matrix and $\|e\|_2
\leq \epsilon$. Contrary to $\ell_0$ minimization, \eqref{eq:L1_min},
which is a convex program, can be solved efficiently. Consequently, it
is possible to recover a stable and robust approximation of $x$ by
solving \eqref{eq:L1_min} instead of \eqref{eq:L0_min} at the cost of
increasing the number of measurements taken.

Several works in the literature have proposed alternate algorithms
that attempt to bridge the gap between $\ell_0$ and $\ell_1$
minimization. For example, the recovery from compressed sensing
measurements using $\ell_p$ minimization with $0 < p < 1$ has been
shown to be stable and robust under weaker conditions that those of
$\ell_1$ minimization \cite{gribonval07:_highl, Foucart08,
  saab2008ssa,chartrand2008rip,
  Saab_ellp:2010}. However, the
problem is non-convex and even though various simple and efficient
algorithms were proposed and observed to perform well empirically
\cite{chartrand2007,saab2008ssa}, so far only local convergence can be
proved. Another approach for improving the recovery performance of
$\ell_1$ minimization is to incorporate prior knowledge regarding the
support of the signal to-be-recovered. One way to accomplish this is
to replace $\ell_1$ minimization in \eqref{eq:L1_min} with
\emph{weighted $\ell_1$
  minimization}
\begin{equation}\label{eq:weighted_L1_0}
  \minim_{u}\ \|u\|_{1,\mathrm{w}}\ \text{subject to}\ \|Au - y\|_2 \leq \epsilon,
\end{equation}
where $\mathrm{w}\in [0,1]^N$ and $\|u\|_{1,\mathrm{w}} := \sum_i
\mathrm{w}_i |u_i|$ is the weighted $\ell_1$ norm. This approach has
been studied by several groups \cite{CS_using_PI_Borries:2007,
  Vaswani_Lu_Modified-CS:2010, Jacques:2010, WL1_min_Hassibi:2009} and
most recently, by the authors, together with Saab and Friedlander
[\citenum{Friedlander_etal:2011}]. In this work, we proved that
conditioned on the accuracy and relative size of the support estimate,
weighted $\ell_1$ minimization is stable and robust under weaker
conditions than those of standard $\ell_1$ minimization.


The works mentioned above mainly focus on a ``two-weight'' scenario:
for $x\in \R^N$, one is given a partition of $\{1,\dots,N\}$ into two
sets, say $\widetilde{T}$ and $\widetilde{T}^c$. Here $\widetilde{T}$
denotes the estimated support of the entries of $x$ that are largest in magnitude. 
In this paper,
we consider the more general case and study recovery conditions of
weighted $\ell_1$ minimization when multiple support estimates with
different accuracies are available. We first give a brief overview of
compressed sensing and review our previous result on weighted $\ell_1$
minimization in Section \ref{sec:2set_wL1}. In Section
\ref{sec:multiset_wL1}, we prove that for a certain class of signals
it is possible to estimate the support of its best $k$-term
approximation using standard $\ell_1$ minimization. We then derive
stability and robustness guarantees for weighted $\ell_1$ minimization
which generalizes our previous work to the case of two or more
weighting sets. Finally, we present numerical experiments in Section
\ref{sec:NumericalResults} that verify our theoretical results.

\section{Compressed sensing with partial support information}\label{sec:2set_wL1} 
Consider an arbitrary signal $x \in \R^N$ and let $x_k$ be its best
$k$-term approximation, given by keeping the $k$ largest-in-magnitude
components of $x$ and setting the remaining components to zero. Let
$T_0 = \supp(x_k)$, where $T_0\subseteq \{1,\dots,N\}$ and $|T_0|
\leq k$. We wish to reconstruct the signal $x$ from $y = Ax + e$,
where $A$ is a known $n\times N$ measurement matrix with $n \ll N$,
and $e$ denotes the (unknown) measurement error that satisfies
$\|e\|_2 \leq \epsilon$ for some known margin $\epsilon>0$. Also let
the set $\widetilde{T} \subset \{1,\dots,N\}$ be an estimate of the
support $T_0$ of $x_k$.

\subsection{Compressed sensing overview} It was shown in
[\citenum{CRT05}] that $x$ can be stably and robustly recovered from
the measurements $y$ by solving the optimization problem
\eqref{eq:L0_min} if the measurement matrix $A$ has the {\em
  restricted isometry property \cite{CandesTao05_2}} (RIP).
\begin{defn}\label{def:RIP}
  The restricted isometry constant $\delta_k$ of a matrix $A$ is the
  smallest number such that for all $k$-sparse vectors
  $u$,
  \begin{equation}\label{eq:RIP} (1-\delta_k)\|u\|_2^2 \leq \|Au\|_2^2
    \leq (1+\delta_k)\|u\|_2^2. \end{equation} \
\end{defn} 
\noindent
The following theorem uses the RIP to provide conditions and
bounds for stable and robust recovery of $x$ by solving
(\ref{eq:L1_min}).\\
\medskip
\begin{theorembf}[Cand$\grave{\textrm{e}}$s, Romberg,
  Tao \cite{CRT05}]\label{thm:L1_recovery}
  Suppose that $x$ is an arbitrary vector in $\R^N$, and let $x_k$ be
  the best $k$-term approximation of $x$. Suppose that there exists an
  $a \in \frac{1}{k}\Z$ with $a>1$ and
  \begin{equation}\label{eq:suff_L1}
    \delta_{ak} + a\delta_{(1+a)k} < a-1.
  \end{equation}
  Then the solution $x^*$ to
  (\ref{eq:L1_min}) obeys
	\begin{equation}\label{eq:L1_recovery}
          \|x^*-x\|_2 \leq C_0\epsilon + C_1 k^{-1/2}\|x - x_k\|_1.
	\end{equation}
\end{theorembf}

\begin{rem}
The constants in Theorem \ref{thm:L1_recovery} are explicitly given by
\large{
\begin{equation}\label{eq:L1_constants}
\begin{array}{l}
	C_0 = \frac{2\left(1+a^{-1/2}\right)}{\sqrt{1-\delta_{(a+1)k}} - a^{-1/2}\sqrt{1+\delta_{ak}}}, \quad C_1 = \frac{2 a^{-1/2}\left(\sqrt{1-\delta_{(a+1)k}} + \sqrt{1+\delta_{ak}}\right) }{\sqrt{1-\delta_{(a+1)k}} - a^{-1/2}\sqrt{1+\delta_{ak}}}.
\end{array}
\end{equation}
}
\normalsize
\end{rem}

Theorem \ref{thm:L1_recovery} shows that the constrained $\ell_1$
minimization problem in (\ref{eq:L1_min}) recovers an approximation to
$x$ with an error that scales well with noise and the
``compressibility'' of $x$, provided \eqref{eq:suff_L1} is satisfied.
Moreover, if $x$ is sufficiently sparse (i.e., $x = x_k$), and if the
measurement process is noise-free, then Theorem \ref{thm:L1_recovery}
guarantees exact recovery of $x$ from $y$. At this point, we note that
a slightly stronger sufficient condition compared to
\eqref{eq:suff_L1}---that is easier to compare with conditions we
obtain in the next section---is given by
\begin{equation}\label{eq:L1_RIP} \delta_{(a+1)k} < \frac{a-1}{a+1}.
\end{equation}

\subsection{Weighted $\ell_1$ minimization} 
\label{sec:wL1}
The $\ell_1$ minimization problem \eqref{eq:L1_min} does not
incorporate any prior information about the support of $x$. However,
in many applications it may be possible to draw an estimate of the
support of the signal or an estimate of the indices of its largest
coefficients.

In our previous work\cite{Friedlander_etal:2011}, we considered the
case where we are given a support estimate $\widetilde{T} \subset
\{1,\dots,N\}$ for $x$ with a certain accuracy. We investigated the
performance of weighted $\ell_1$ minimization, as described in
\eqref{eq:weighted_L1_0}, where the weights are assigned such that
$\mathrm{w}_j=\omega \in [0,1]$ whenever $j\in \widetilde{T}$, and
$\mathrm{w}_j=1$ otherwise. In particular, we proved that if the
(partial) support estimate is \textit{at least} 50\% accurate, then
weighted $\ell_1$ minimization with $\omega<1$ outperforms standard
$\ell_1$ minimization in terms of accuracy, stability, and robustness.

\begin{figure}[h]
	\centering
		\includegraphics[width=5.5in]{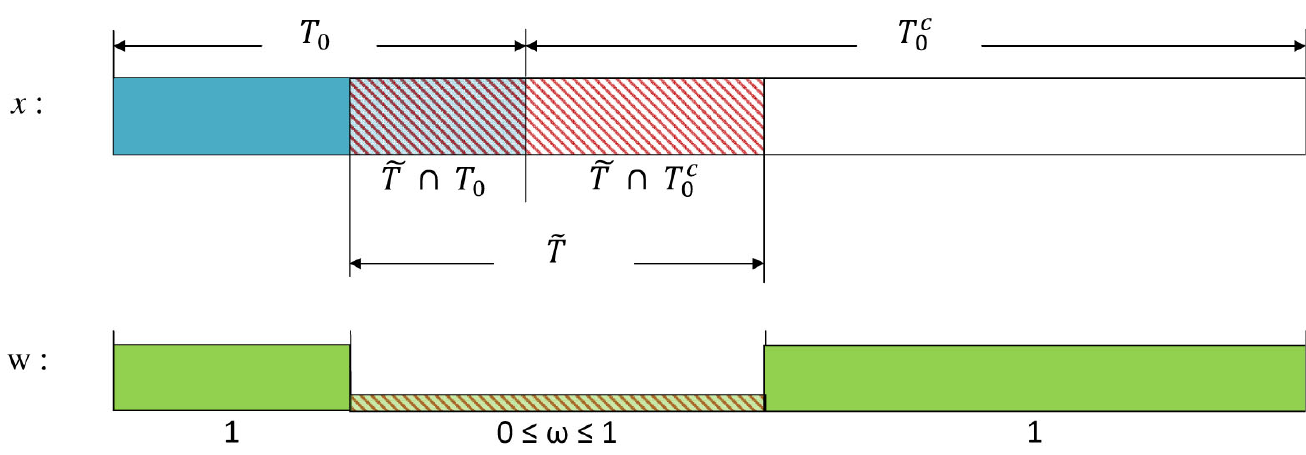}
	\caption{Illustration of the signal $x$ and weight vector $\mathrm{w}$ emphasizing the relationship between the sets $T_0$ and $\widetilde{T}$.  }
	\label{fig:Sets_Weights}
\end{figure}

Suppose that $\widetilde{T}$ has cardinality $|\widetilde{T}| = \rho
k$, where $0 \leq \rho \leq N/k$ is the \emph{relative size} of the
support estimate $\widetilde{T}$. Furthermore, define the
\emph{accuracy} of $\widetilde{T}$ via $\alpha:=
\frac{\widetilde{T}\cap
  T_0}{|\widetilde{T}|}$, i.e., $\alpha$ is the fraction of
$\widetilde{T}$ inside $T_0$. As before, we wish to recover an
arbitrary vector $x \in \R^N$ from noisy compressive measurements $y =
Ax + e$, where $e$ satisfies $\|e\|_2 \leq \epsilon$. To that end, we
consider the weighted $\ell_1$ minimization problem with the following
choice of weights: \begin{equation}\label{eq:weighted_L1}
  \minim_{z}\ \|z\|_{1,\mathrm{w}}\ \text{subject to}\ \|Az - y\|_2 \leq \epsilon
  \quad\text{with}\quad
  \mathrm{w}_i = 
  \begin{cases}
      1,      & i \in \widetilde{T}^c,
    \\\omega, & i \in \widetilde{T}. 
  \end{cases}
\end{equation}
Here, $0 \leq \omega \leq 1$ and $\|z\|_{1,\mathrm{w}}$ is as defined
in (\ref{eq:weighted_L1_0}). Figure \ref{fig:Sets_Weights} illustrates
the relationship between the support $T_0$, support estimate
$\widetilde{T}$ and the weight vector $\mathrm{w}$.
\medskip

\noindent
\begin{theorembf}[FMSY \cite{Friedlander_etal:2011}]\label{thm:weighted_L1_recovery}
  Let $x$ be in $\R^N$ and let $x_k$ be its best $k$-term
  approximation, supported on $T_0$.  Let
  $\widetilde{T}\subset\{1,\ldots,N\}$ be an arbitrary set and define
  $\rho$ and $\alpha$ as before such that $|\widetilde{T}|=\rho k$ and
  $|\widetilde{T} \cap T_0| = \alpha\rho k$.  Suppose that there
  exists an $a\in \frac{1}{k}\Z$, with $a \geq (1-\alpha)\rho$, $a>1$,
  and the measurement matrix $A$ has RIP with
\begin{equation}\label{eq:suff_wl1}
\delta_{ak} + \frac{a}{\left(\omega + (1-\omega)\sqrt{1+\rho-2\alpha\rho}\right)^2}\delta_{(a+1)k} < \frac{a}{\left(\omega + (1-\omega)\sqrt{1+\rho-2\alpha\rho}\right)^2} - 1, 
\end{equation}
for some given $0 \leq \omega \leq 1$. Then the solution $x^*$ to (\ref{eq:weighted_L1}) obeys
\begin{equation}\label{eq:weighted_L1_recovery}
	\|x^* - x\|_2 \leq C_0'\epsilon + C_1'k^{-1/2}\left(\omega\|x - x_k\|_1 + (1-\omega)\|x_{\widetilde{T}^c\cap T_0^c}\|_1\right),
\end{equation}
where $C_0'$ and $C_1'$ are well-behaved constants that depend on the measurement matrix $A$, the weight $\omega$, and the parameters $\alpha$ and $\rho$.
\end{theorembf}
 
\begin{rem}\label{rem:const}
The constants $C_0'$ and $C_1'$ are explicitly given by the expressions
\begin{equation}\label{eq:weighted_L1_constants}
\begin{array}{ll}
	C_0' = \frac{\textstyle 2\left(1+\frac{\omega +
              (1-\omega)\sqrt{1+\rho-2\alpha\rho}}{\sqrt{a}}\right)}{\textstyle\sqrt{1-\delta_{(a+1)k}}
          - \frac{\omega +
            (1-\omega)\sqrt{1+\rho-2\alpha\rho}}{\sqrt{a}}\sqrt{1+\delta_{ak}}},&
        C_1' = \frac{\textstyle 2
          a^{-1/2}\left(\sqrt{1-\delta_{(a+1)k}} +
            \sqrt{1+\delta_{ak}}\right) }{\textstyle \sqrt{1-\delta_{(a+1)k}} - \frac{\omega + (1-\omega)\sqrt{1+\rho-2\alpha\rho}}{\sqrt{a}}\sqrt{1+\delta_{ak}}}.
\end{array}
\end{equation}

\noindent Consequently, Theorem~\ref{thm:weighted_L1_recovery}, with
$\omega=1$, reduces to the stable and robust recovery theorem of
[\citenum{CRT05}], which we stated above---see
Theorem~\ref{thm:L1_recovery}.
\end{rem}

\begin{rem} It is sufficient that $A$ satisfies
\begin{equation}\label{eq:weighted_L1_RIP}
	\delta_{(a+1)k} < \hat{\delta}^{(\omega)} := \frac{a - \left(\omega + (1-\omega)\sqrt{1+\rho - 2\alpha\rho}\right)^2}{a + \left(\omega + (1-\omega)\sqrt{1+\rho - 			2\alpha\rho}\right)^2}
\end{equation}
for Theorem \ref{thm:weighted_L1_recovery} to hold, i.e., to guarantee
stable and robust recovery of the signal $x$ from measurements $y = Ax
+ e$.
\end{rem}

It is easy to see that the sufficient conditions of
Theorem~\ref{thm:weighted_L1_recovery}, given in \eqref{eq:suff_wl1}
or \eqref{eq:weighted_L1_RIP}, are weaker than their counterparts for
the standard $\ell_1$ recovery, as given in \eqref{eq:suff_L1} or
\eqref{eq:L1_RIP} respectively, if and only if $\alpha>0.5$. A similar
statement holds for the constants. In words, if the support estimate
is more than 50\% accurate, weighted $\ell_1$ is more favorable than
$\ell_1$, at least in terms of sufficient conditions and error bounds.

The theoretical results presented above suggest that the weight
$\omega$ should be set equal to zero when $\alpha \geq 0.5$ and to one
when $\alpha < 0.5$ as these values of $\omega$ give the best
sufficient conditions and error bound constants. However, we conducted
extensive numerical simulations in [\citenum{Friedlander_etal:2011}]
which suggest that a choice of $\omega \approx 0.5$ results in the
best recovery when there is little confidence in the support estimate
accuracy. An heuristic explanation of this observation is given in
[\citenum{Friedlander_etal:2011}].

\section{Weighted $\ell_1$ minimization with multiple support estimates}\label{sec:multiset_wL1} 
The result in the previous section relies on the availability of a
support estimate set $\widetilde{T}$ on which to apply the weights
$\omega$. In this section, we first show that it is possible to draw
support estimates from the solution of \eqref{eq:L1_min}. We then
present the main theorem for stable and robust recovery of an
arbitrary vector $x \in \R^N$ from measurements $y = Ax + e$, $y \in
\R^n$ and $n \ll N$, with multiple support estimates having different
accuracies. 

\subsection{Partial support recovery from $\ell_1$ minimization} 
For signals $x$ that belong to certain signal classes, the solution to
the $\ell_1$ minimization problem can carry significant information on
the support $T_0$ of the best $k$-term approximation $x_k$ of $x$. We
start by recalling the \textit{null space property} (NSP) of a matrix
$A$ as defined in [\citenum{cohen2006csa}]. Necessary conditions as
well as sufficient conditions for the existence of some algorithm that
recovers $x$ from measurements $y = Ax$ with an error related to the
best $k$-term approximation of $x$ can be formulated in terms of an
appropriate NSP. We state below a particular form of the NSP
pertaining to the $\ell_1$-$\ell_1$ instance optimality.

\begin{defn}\label{def:NSP}
  A matrix $A\in \R^{n\times N}$, $n < N$, is said to have the null
  space property of order $k$ and constant $c_0$ if for any vector $h
  \in \mathcal{N}(A)$, $Ah = 0$, and for every index set $T \subset
  \{1\dots N\}$ of cardinality $|T| = k$
	$$
		\|h\|_1 \leq c_0 \|h_{T^c}\|_1.
	$$
\end{defn}

Among the various important conclusions of [\citenum{cohen2006csa}],
the following (in a slightly more general form) will be instrumental
for our results.

\begin{lemmabf}[{[\citenum{cohen2006csa}]}] If $A$ has the restricted isometry
  property with $\delta_{(a+1)k} < \frac{a-1}{a+1}$ for some $a>1$,
  then it has the NSP of order $k$ and constant $c_0$ given explicitly
  by
	$$
		c_0 = 1 + \frac{\sqrt{1 + \delta_{ak}}}{\sqrt{a}\sqrt{1 - \delta_{(a+1)k}}}.
	$$
\end{lemmabf}

In what follows, let $x^*$ be the solution to \eqref{eq:L1_min} and
define the sets $S = \textrm{supp}(x_{s})$, $T_0 =
\textrm{supp}(x_{k})$, and $\widetilde{T} = \textrm{supp}(x^*_{k})$
for some integers $k \ge s > 0$.

\begin{propbf}\label{lemma:partSupportL1}
  Suppose that $A$ has the null space property (NSP) of order $k$ with
  constant $c_0$ and
	\begin{equation}\label{eq:partSupportL1}
		\min_{j \in S}|x(j)| \geq (\eta + 1)\|x_{T_0^c}\|_1,
	\end{equation}
where $\eta = \frac{2c_0}{2 - c_0}$. Then $S \subseteq \widetilde{T}$.
\end{propbf}

The proof is presented in section \ref{sec:proofLemma} of the
appendix.

\begin{rem} Note that if $A$ has RIP so that $\delta_{(a+1)k} <
  \frac{a-1}{a+1}$ for some $a>1$, then $\eta$ is given explicitly by
  \begin{equation}\label{eq:eta} \eta = \frac{2(\sqrt{a}\sqrt{1 -
        \delta_{(a+1)k}} + \sqrt{1+\delta_{ak}})}{\sqrt{a}\sqrt{1 -
        \delta_{(a+1)k}} - \sqrt{1+\delta_{ak}}}. \end{equation}
\end{rem}

Proposition \ref{lemma:partSupportL1} states that if $x$ belongs to
the class of signals that satisfy \eqref{eq:partSupportL1}, then the
support $S$ of $x_s$---i.e., the set of indices of the $s$
largest-in-magnitude coefficients of $x$---is guaranteed to be
contained in the set of indices of the $k$ largest-in-magnitude
coefficients of $x^*$. Consequently, if we consider $\widetilde{T}$ to
be a support estimate for $x_k$, then it has an accuracy $\alpha \geq
\frac{s}{k}$.

Note here that Proposition \ref{lemma:partSupportL1} specifies a class
of signals, defined via \eqref{eq:partSupportL1}, for which partial
support information can be obtained by using the standard $\ell_1$
recovery method. Though this class is quite restrictive and does not
include various signals of practical interest, experiments suggest
that highly accurate support estimates can still be obtained via
$\ell_1$ minimization for signals that only satisfy significantly
milder decay conditions than \eqref{eq:partSupportL1}. A theoretical
investigation of this observation is an open problem.

\subsection{Multiple support estimates with varying accuracy: an
  idealized motivating example} 
Suppose that the entries of $x$ decay according to a power law such
that $|x(j)| = cj^{-p}$ for some scaling constant $c$, $p >1$ and $j
\in \{1,\dots, N\}$. Consider the two support sets ${T}_1 =
\textrm{supp}(x_{k_1})$ and ${T}_2 = \textrm{supp}(x_{k_2})$ for $k_1
> k_2$, $T_2 \subset T_1$. Suppose also that we can find entries
$|x(s_1)| = c s_1^{-p} \approx c(\eta + 1)\frac{k_1^{1-p}}{p-1}$ and
$x(s_2) = c s_2^{-p} \approx c(\eta + 1)\frac{k_2^{1-p}}{p-1}$ that
satisfy \eqref{eq:partSupportL1} for the sets $T_1$ and $T_2$,
respectively, where $s_1 \leq k_1$ and $s_2 \leq k_2$. Then 

$$
\begin{array}{ll}
s_1 - s_2 &= \left(\frac{p-1}{\eta+1}\right)^{1/p}\left(k_1^{1-1/p} - k_2^{1-1/p}\right) \\
& \le \left(\frac{p-1}{\eta+1}\right)^{1/p}(k_1 - k_2).
\end{array}
$$
which follows because $0<1-1/p<1$ and $k_1-k_2 \ge 1$.

Consequently, if we define the support estimate sets $\widetilde{T}_1
= \textrm{supp}(x^*_{k_1})$ and $\widetilde{T}_2 =
\textrm{supp}(x^*_{k_2})$, clearly the corresponding accuracies
$\alpha_1 = \frac{s_1}{k_1}$ and $\alpha_2 = \frac{s_2}{k_2}$ are not
necessarily equal. Moreover, if
\begin{equation}\label{eq:order_relation}  
  \left(\frac{p-1}{\eta+1}\right)^{1/p} < \alpha_1, 
\end{equation}
$s_1 - s_2 < \alpha_1(k_1 - k2)$, and thus $\alpha_1 < \alpha_2$. For
example, if we have $p=1.3$ and $\eta=5$, we get
$\left(\frac{p-1}{\eta+1}\right)^{1/p} \approx 0.1$. Therefore, in
this particular case, if $\alpha_1 > 0.1$, choosing some $k_2 < k_1$
results in $\alpha_2 > \alpha_1$, i.e., we identify two different
support estimates with different accuracies. This observation raises
the question, ``How should we deal with the recovery of signals from
CS measurements when multiple support estimates with different
accuracies are available?''  We propose an answer to this question in
the next section.

\subsection{Stability and robustness conditions} 
\begin{figure*}[h]
	\centering
	\includegraphics[width=5in, height=3.5in]{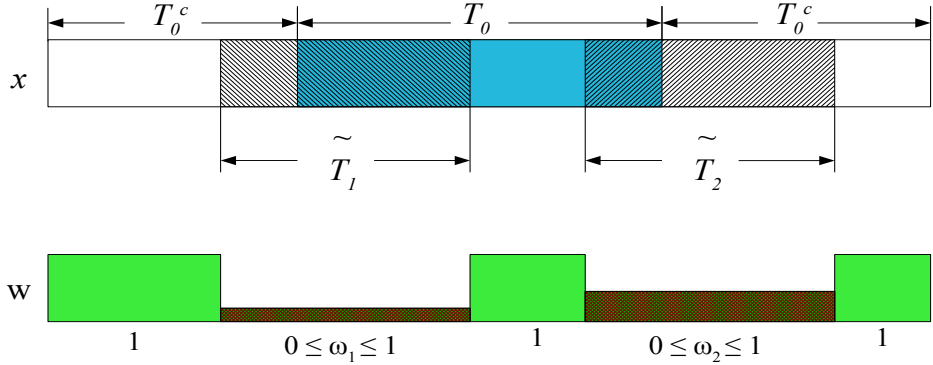} 
	\caption{Example of a sparse vector $x$ with support set $T_0$
          and two support estimate sets $\widetilde{T}_1$ and
          $\widetilde{T}_2$. The weight vector is chosen so that
          weights $\omega_1$ and $\omega_2$ are applied to the sets
          $\widetilde{T}_1$ and $\widetilde{T}_2$, respectively, and a
          weight equal to one elsewhere.}
	\label{fig:3set_weight}
\end{figure*}

In this section we present our main theorem for stable and robust
recovery of an arbitrary vector $x \in \R^N$ from measurements $y = Ax
+ e$, $y \in \R^n$ and $n \ll N$, with multiple support estimates
having different accuracies. Figure \ref{fig:3set_weight} illustrates
an example of the particular case when only two disjoint support
estimate sets are available.

Let $T_0$ be the support of the best $k$-term approximation $x_k$ of
the signal $x$. Suppose that we have a support estimate $\widetilde{T}$
that can be written as the union of $m$ disjoint subsets
$\widetilde{T}_j$, $j \in \{1,\dots,m\}$, each of which has
cardinality $|\widetilde{T}_j| = \rho_j k$, $0 \leq \rho_j \leq a$ for
some $a > 1$ and accuracy $\alpha_j = \frac{|\widetilde{T}_j\cap
  T_0|}{|\widetilde{T}_j|}$.

Again, we wish to recover $x$ from measurements $y = Ax + e$ with
$\|e\|_2 \leq \epsilon$. To do this, we consider the general weighted
$\ell_1$ minimization problem
\begin{equation}\label{eq:multiset_wL1_min}
	\min\limits_{u \in \R^N} \|u\|_{1,\mathrm{w}} \quad \textrm{subject to} \quad \|Au - y\| \leq \epsilon
\end{equation}
where $\|u\|_{1,\mathrm{w}} = \sum\limits_{i=1}^N \mathrm{w}_i|u_i|$, and 
$\mathrm{w}_i = \left\{
\begin{array}{ll}
	\omega_1,&  i \in \widetilde{T}_1\\
	\vdots&\\
	\omega_m,&  i \in \widetilde{T}_m\\
	1, & i \in \widetilde{T}^c  
\end{array}\right.$ for $0 \leq \omega_j \leq 1$, for all $j \in
\{1,\dots, m\}$ and $\widetilde{T} = \bigcup\limits_{j =
  1}^m\widetilde{T}_j$. 
\medskip

\noindent
\begin{theorembf}\label{thm:multiset_wL1}
  Let $x \in \R^n$ and $y=Ax+e$, where $A$ is an $n\times N$ matrix
  and $e$ is additive noise with $\|e\|_2 \le \epsilon$ for some known
  $\epsilon>0$. Denote by $x_k$ the best $k$-term approximation of
  $x$, supported on $T_0$ and let $\widetilde{T}_1,\dots,
  \widetilde{T}_m \subset\{1,...,N\}$ be as defined above with
  cardinality $|\widetilde{T}_j| = \rho_j k$ and accuracy $\alpha_j =
  \frac{|\widetilde{T}_j\cap T_0|}{|\widetilde{T}_j|}$, $j \in
  \{1,\dots, m\}$. For some given $0 \leq \omega_1,\dots, \omega_m
  \leq 1$, define $\gamma := \sum\limits_{j = 1}^m\omega_j - (m-1) +
  \sum\limits_{j=1}^m(1-\omega_j)\sqrt{1+\rho_j-2\alpha_j\rho_j}$. If
  the RIP constants of $A$ are such that there exists an $a\in
  \frac{1}{k}\Z$, with $a>1$, and
\begin{equation}\label{eq:suff_multiset_wl1}
\delta_{ak} + \frac{a}{\gamma^2}\delta_{(a+1)k} < \frac{a}{\gamma^2} - 1, 
\end{equation}
then the solution $x^\#$ to (\ref{eq:multiset_wL1_min}) obeys
\begin{equation}\label{eq:weighted_L1_recovery}
	\|x^\# - x\|_2 \leq C_0(\gamma)\epsilon + C_1(\gamma)k^{-1/2}\left(\sum\limits_{j = 1}^m\omega_j\|x_{\widetilde{T}_j\cap T_0^c}\|_1 + \|x_{\widetilde{T}^c\cap T_0^c}\|_1\right).
\end{equation}
\end{theorembf} 
The proof is presented in section \ref{sec:proofTheorem} of the appendix.

\begin{rem}\label{rem:const_multiset}
The constants $C_0(\gamma)$ and $C_1(\gamma)$ are well-behaved and given explicitly by the expressions
\begin{equation}\label{eq:weighted_L1_constants}
	C_0(\gamma) = \frac{\textstyle 2\left(1+\frac{\gamma}{\sqrt{a}}\right)}{\textstyle\sqrt{1-\delta_{(a+1)k}}
          - \frac{\gamma}{\sqrt{a}}\sqrt{1+\delta_{ak}}},\quad
        C_1(\gamma) = \frac{\textstyle 2
          a^{-1/2}\left(\sqrt{1-\delta_{(a+1)k}} +
            \sqrt{1+\delta_{ak}}\right) }{\textstyle \sqrt{1-\delta_{(a+1)k}} - \frac{\gamma}{\sqrt{a}}\sqrt{1+\delta_{ak}}}.
\end{equation}
\end{rem}

\begin{rem}
Theorem \ref {thm:multiset_wL1} is a generalization of Theorem \ref{thm:weighted_L1_recovery} for $m \geq 1$ support estimates. It is easy to see that when the number of support estimates $m = 1$, Theorem \ref {thm:multiset_wL1} reduces to the recovery conditions of Theorem \ref{thm:weighted_L1_recovery}. Moreover, setting $\omega_j = 1$ for all $j \in \{1,\dots, m\}$ reduces the result to that in Theorem \ref{thm:L1_recovery}. 
\end{rem}

\begin{rem}
The sufficient recovery condition \eqref{eq:weighted_L1_RIP} becomes in the case of multiple support estimates
\begin{equation}\label{eq:multiset_wL1_RIP}
	\delta_{(a+1)k} < \hat{\delta}^{(\gamma)} := \frac{a - \gamma^2}{a + \gamma^2},
\end{equation}
where $\gamma$ is as defined in Theorem~\ref{thm:multiset_wL1}.
It can be shown that when $m = 1$, $\gamma$ reduces to the expression in \eqref{eq:weighted_L1_RIP}. 
\end{rem}

\begin{rem}
  The value of $\gamma$ controls the recovery guarantees of the
  multiple-set weighted $\ell_1$ minimization problem. For instance,
  as $\gamma$ approaches 0, condition \eqref{eq:multiset_wL1_RIP}
  becomes weaker and the error bound constants $C_0(\gamma)$ and
  $C_1(\gamma)$ become smaller. Therefore, given a set of support
  estimate accuracies $\alpha_j$ for all $j \in \{1\dots m\}$, it is
  useful to find the corresponding weights $\omega_j$ that minimize
  $\gamma$. Notice that for all $j$, $\gamma$ is a sum of linear
  functions of $\omega_j$ with $\alpha_j$ controlling the slope. When
  $\alpha_j > 0.5$, the slope is positive and the optimal value of
  $\omega_j = 0$. Otherwise, when $\alpha_j \leq 0.5$, the slope is
  negative and the optimal value of $\omega_j = 1$. Hence, as in the
  single support estimate case, the theoretical conditions indicate
  that when the $\alpha_j$ are known a choice of $\omega_j$ equal to
  zero or one should be optimal. However, when the knowledge of
  $\alpha_j$ is not reliable, experimental results indicate that
  intermediate values of $\omega_j$ produce the best recovery results.
\end{rem}

\section{Numerical experiments} \label{sec:NumericalResults} In what
follows, we consider the particular case of $m = 2$, i.e. where there
exists prior information on two disjoint support estimates
$\widetilde{T}_1$ and $\widetilde{T}_2$ with respective accuracies
$\alpha_1$ and $\alpha_2$. We present numerical experiments that
illustrate the benefits of using three-set weighted $\ell_1$
minimization over two-set weighted $\ell_1$ and non-weighted $\ell_1$
minimization when additional prior support information is available.

To that end, we compare the recovery capabilities of these algorithms
for a suite of synthetically generated sparse signals. We also present
the recovery results for a practical application of recovering audio
signals using the proposed weighting. In all of our experiments, we
use SPGL1~\cite{BergFriedlander:2008, spgl1:2007} to solve the
standard and weighted $\ell_1$ minimization problems.

\subsection{Recovery of synthetic signals}
We generate signals $x$ with an ambient dimension $N = 500$ and fixed
sparsity $k = 35$. We compute the (noisy) compressed measurements of
$x$ using a Gaussian random measurement matrix $A$ with dimensions $n
\times N$ where $n=100$. To quantify the reconstruction quality,
we use the reconstruction signal to noise ratio (SNR) average over 100
realizations of the same experimental conditions. The SNR is measured
in dB and is given by
\begin{equation}
	\mathrm{SNR}(x,\tilde{x}) = 10\log_{10}\left(\frac{\|x\|_2^2}{\|x - \tilde{x}\|_2^2}\right),
\end{equation}
where $x$ is the true signal and $\tilde{x}$ is the recovered signal. 

The recovery via two-set weighted $\ell_1$ minimization uses a support
estimate $\widetilde{T}$ of size $|\widetilde{T}| = 40$ (i.e., $\rho =
1$) where the accuracy $\alpha$ of the support estimate takes on the
values $\{0.3, 0.5, 0.7\}$, and the weight $\omega$ is chosen from
$\{0.1, 0.3, 0.5\}$.

Recovery via three-set weighted $\ell_1$ minimization assumes the existence of two support estimates $\widetilde{T}_1$ and $\widetilde{T}_2$, which are disjoint subsets of $\widetilde{T}$ described above. The set $\widetilde{T}_1$ is chosen such that it always has an accuracy $\alpha_1 = 0.8$ while $\widetilde{T}_2 = \widetilde{T}\setminus \widetilde{T}_1$. In all experiments, we fix $\omega_1 = 0.01$ and set $\omega_2 = \omega$. 

\begin{figure*}\label{fig:2set_vs_3set_wL1}
	\centering
	\includegraphics[width = 7.5in]{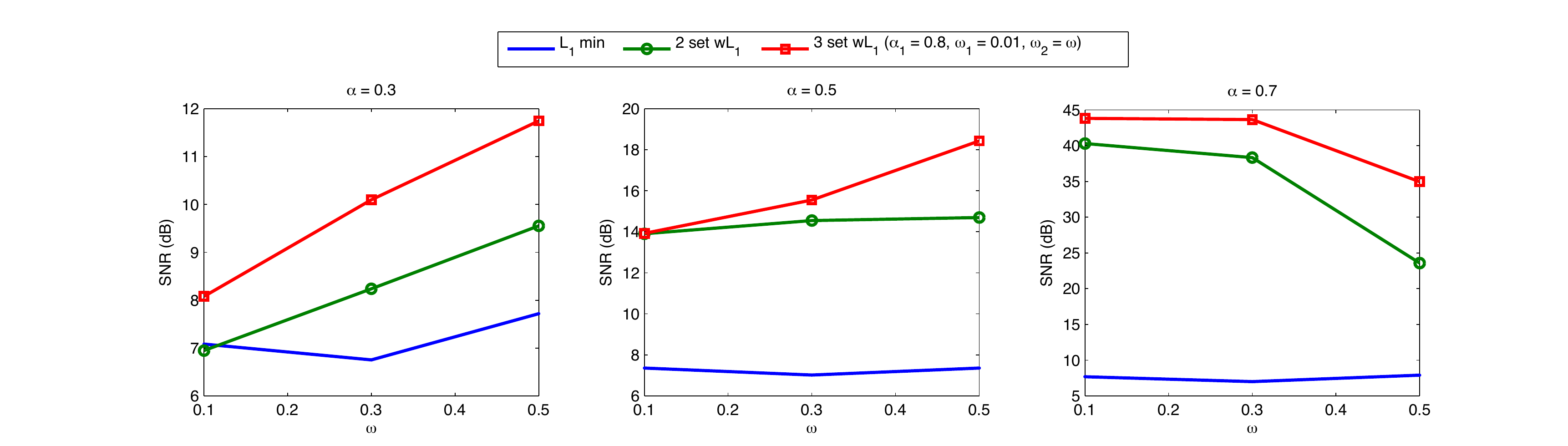}
	\caption{Comparison between the recovered SNR (averaged over 100 experiments) using two-set weighted $\ell_1$ with support estimate $\widetilde{T}$ and accuracy $\alpha$, three-set weighted $\ell_1$ minimization with support estimates $\widetilde{T}_1 \cup \widetilde{T}_2 = \widetilde{T}$ and accuracy $\alpha_1 = 0.8$ and $\alpha_2 < \alpha$, and non-weighted $\ell_1$ minimization.}
\end{figure*}

Figure \ref{fig:2set_vs_3set_wL1} illustrates the recovery performance of three-set weighted $\ell_1$ minimization compared to two-set weighted $\ell_1$ using the setup described above and non-weighted $\ell_1$ minimization. The figure shows that utilizing the extra accuracy of $\widetilde{T}_1$ by setting a smaller weight $\omega_1$ results in better signal recovery from the same measurements.

\subsection{Recovery of audio signals}
Next, we examine the performance of three-set weighted $\ell_1$
minimization for the recovery of compressed sensing measurements of
speech signals. In particular, the original signals are sampled at
$44.1$ kHz, but only $1/4$th of the samples are retained (with their
indices chosen randomly from the uniform distribution). This yields
the measurements $y=Rs$, where $s$ is the speech signal and $R$ is a
restriction (of the identity) operator. Consequently, by dividing the
measurements into blocks of size $N$, we can write
$y=[y_1^T,y_2^T,...]^T$. Here each $y_j=R_j s_j$ is the measurement
vector corresponding to the $j$th block of the signal, and
$R_j\in\R^{n_j\times N}$ is the associated restriction matrix. The
signals we use in our experiments consist of 21 such blocks.

\begin{figure*}[h]
	\centering
	\includegraphics[width=5in]{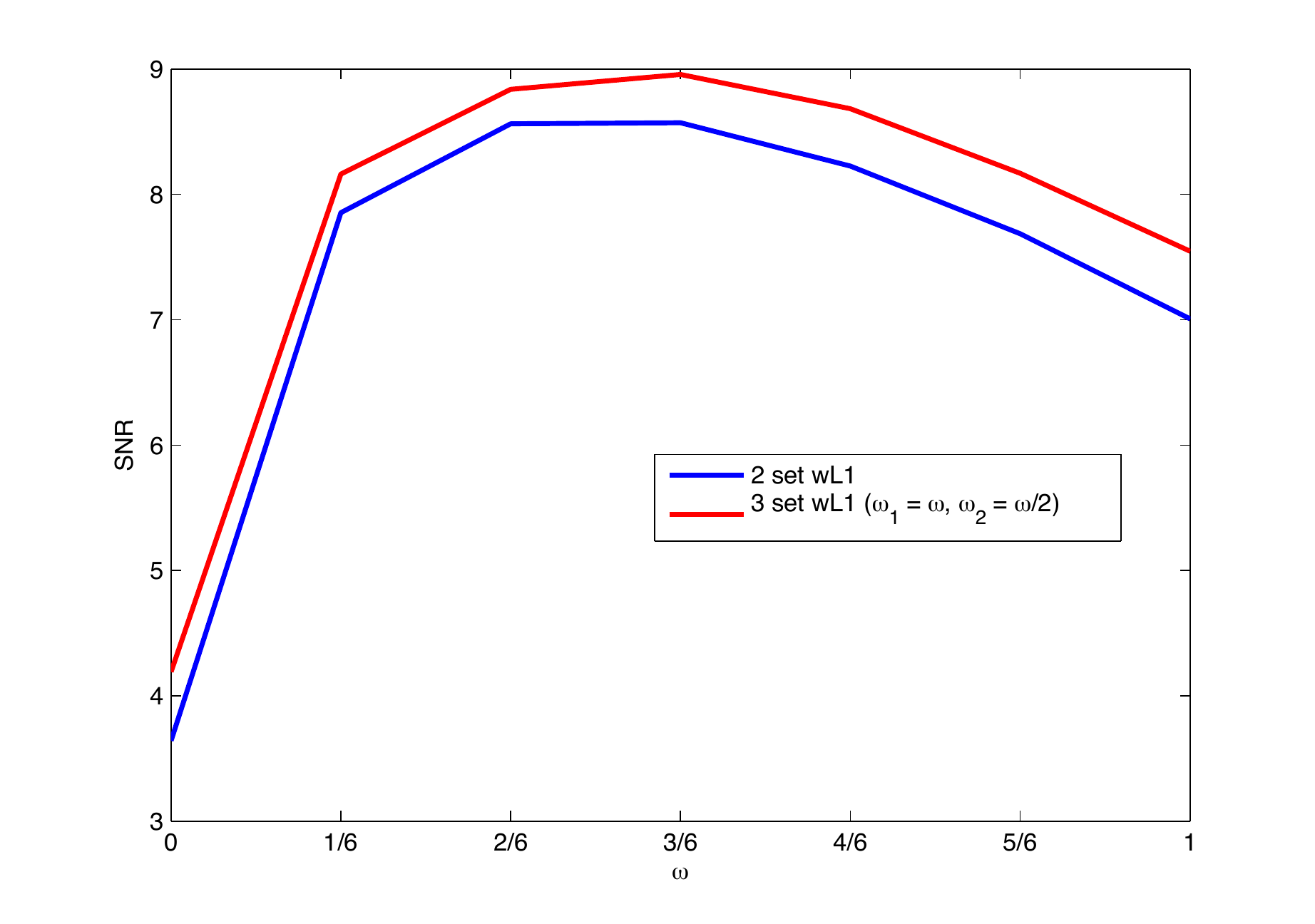}
	\caption{SNRs of the two reconstruction algorithms two-set and
          three-set weighted $\ell_1$ minimization for a speech signal
          from compressed sensing measurements plotted against
          $\omega$.}
	\label{fig:audio}
\end{figure*}

We make the following assumptions about speech signals: 
\begin{enumerate}
\item The signal blocks are compressible in the DCT domain (for example, the MP3
compression standard uses a version of the  DCT to compress audio
signals.)
\item The support set corresponding to the largest coefficients in
  adjacent blocks does not change much from block to block.
\item Speech signals have large low-frequency coefficients.
\end{enumerate}
Thus, for the reconstruction of the $j$th block, we identify the
support estimates $\widetilde{T}_1$ is the set corresponding to the
largest $n_j/16$ recovered coefficients of the previous block (for the
first block $\widetilde{T}_1$ is empty) and $\widetilde{T}_2$ is the
set corresponding to frequencies up to 4kHz. For recovery using
two-set weighted $\ell_1$ minimization, we define
$\widetilde{T}=\widetilde{T}_1\cup\widetilde{T}_2$ and assign it a
weight of $\omega$. In the three-set weighted $\ell_1$ case, we assign
weights $\omega_1 = \omega/2$ on the set $\widetilde{T}_1$ and
$\omega_2 = \omega$ on the set
$\widetilde{T}\setminus\widetilde{T}_1$. The results of experiments on
an example speech signal with $N=2048$, and $\omega \in
\{0,1/6,2/6,\ldots,1\}$ are illustrated in Figure~\ref{fig:audio}. It
is clear from the figure that three-set weighted $\ell_1$ minimization
has better recovery performance over all 10 values of $\omega$
spanning the interval $[0,1]$.
\section{Conclusion} \label{sec:Conclusion} In conclusion, we derived
stability and robustness guarantees for the weighted $\ell_1$
minimization problem with multiple support estimates with varying
accuracy. We showed that incorporating additional support information
by applying a smaller weight to the estimated subsets of the support
with higher accuracy improves the recovery conditions compared with
the case of a single support estimate and the case of (non-weighted)
$\ell_1$ minimization. We also showed that for a certain class of
signals---the coefficients of which decay in a particular way---it is
possible to draw a support estimate from the solution of the $\ell_1$
minimization problem. These results raise the question of whether it
is possible to improve on the support estimate by solving a subsequent
weighted $\ell_1$ minimization problem. Moreover, it raises an
interest in defining a new iterative weighted $\ell_1$ algorithm which
depends on the support accuracy instead of the coefficient magnitude
as is the case of the Cand\`{e}s, Wakin, and Boyd\cite{candes2008irl1}
(IRL1) algorithm. We shall consider these problems elsewhere.

\appendix    
\section{Proof of Proposition 3.2} \label{sec:proofLemma}
We want to find the conditions on the signal $x$ and the matrix $A$ which guarantee that the solution $x^*$ to the $\ell_1$ minimization problem \eqref{eq:L1_min} has the following property
$$
	\min_{j\in S} |x^*(j)| \geq \max_{j\in \widetilde{T}^c} |x^*(j)| = |x^*({k+1})|.
$$

Suppose that the matrix $A$ has the Null Space property (NSP) \cite{cohen2006csa} of order $k$, i.e., for any $h \in \mathcal{N}(A)$, $Ah = 0$, then 
$$
	\|h\|_1 \leq c_0\|h_{T_0^c}\|_1,
$$
where $T_0 \subset \{1,2,\dots N\}$ with $|T_0| = k$, and $\mathcal{N}(A)$ denotes the Null-Space of $A$. 

If $A$ has RIP with $\delta_{(a+1)k} < \frac{a-1}{a+1}$ for some constant $a > 1$, then it has the NSP of order $k$ with constant $c_0$ which can be written explicitly in terms of the RIP constant of $A$ as follows
$$
c_0 = 1 + \frac{\sqrt{1 + \delta_{ak}}}{\sqrt{a}\sqrt{1 - \delta_{(a+1)k}}}.
$$

Define $h = x^* - x$, then $h \in \mathcal{N}(A)$ and we can write the $\ell_1$-$\ell_1$ instance optimality as follows
$$
	\|h\|_1 \leq \frac{2c_0}{2 - c_0}\|x_{T_0^c}\|_1,
$$
with $c_0 < 2$. Let $\eta = \frac{2c_0}{2 - c_0}$, the bound on $\|h_{T_0}\|_1$ is then given by
\begin{equation}\label{eq:hT0}
	\|h_{T_0}\|_1 \leq (\eta + 1)\|x_{T_0^c}\|_1 - \|x^*_{T_0^c}\|_1.
\end{equation}

The next step is to bound $\|x^*_{T_0^c}\|_1$. Noting that $\widetilde{T} = \mathrm{supp}(x^*_{k})$, then $\|x^*_{\widetilde{T}}\|_1 \leq \|x^*_{T_0^c}\|_1$, and 
$$
	\|x^*_{T_0^c}\|_1 \geq \|x^*_{\widetilde{T}^c}\|_1 \geq |x^*({k+1})|.
$$

Using the reverse triangle inequality, we have $\forall j$, $|x(j) - x^*(j)| \geq |x(j)| - |x^*(j)|$ which leads to 
$$
	\min_{j \in S} |x^*(j)| \geq \min_{j \in S}|x(j)| - \max_{j \in S} |x(j) - x^*(j)|.	
$$

But $\max\limits_{j \in S} |x(j) - x^*(j)| = \|h_{S}\|_{\infty} \leq \|h_{S}\|_{1} \leq \|h_{T_0}\|_{1}$, so combining the above three equations we get
\begin{equation}\label{eq:RecoveryCondition}
	\min_{j \in S} |x^*(j)| \geq |x^*({k+1})| + \min_{j \in S}|x(j)| - (\eta + 1)\|x_{T_0^c}\|_1.
\end{equation}
Equation \eqref{eq:RecoveryCondition} says that if the matrix $A$ has $\delta_{(a+1)k}$-RIP and the signal $x$ obeys 
$$
	\min_{j \in S}|x(j)| \geq (\eta + 1)\|x_{T_0^c}\|_1,
$$
then the support $\widetilde{T}$ of the largest $k$ entries of the solution $x^*$ to \eqref{eq:L1_min} contains the support $S$ of the largest $s$ entries of the signal $x$.

\section{Proof of Theorem 3.3} \label{sec:proofTheorem}
The proof of Theorem \ref{thm:multiset_wL1} follows in the same line as our previous work in [\citenum{Friedlander_etal:2011}] with some modifications. Recall that the sets $\widetilde{T}_j$ and disjoint and $\widetilde{T} = \bigcup\limits_{j=1}^m\widetilde{T}_j$, and define the sets $\widetilde{T}_{j\alpha} = T_0 \cap \widetilde{T}_j$, for all $j \in \{1, \dots, m\}$, where $|\widetilde{T}_{j\alpha}| = \alpha_j\rho_jk$.

Let $x^\# = x + h$ be the minimizer of the weighted $\ell_1$ problem \eqref{eq:multiset_wL1_min}. Then
\begin{equation*}
	 \|x + h\|_{1,\mathrm{w}}  \leq \|x\|_{1,\mathrm{w}}.
\end{equation*}
Moreover, by the choice of weights in \eqref{eq:multiset_wL1_min}, we have
\begin{equation*}
 \omega_1\|x_{\widetilde{T}_1} + h_{\widetilde{T}_1}\|_1 + \dots \omega_m\|x_{\widetilde{T}_m} + h_{\widetilde{T}_m}\|_1 + \|x_{\widetilde{T}^c} + h_{\widetilde{T}^c}\|_1 \leq \omega_1\|x_{\widetilde{T}_1}\|_1 \dots +\omega_m\|x_{\widetilde{T}_m}\|_1 + \|x_{\widetilde{T}^c}\|_1. 
\end{equation*}
Consequently, 
\begin{equation*}
\begin{array}{ll}
 & \|x_{\widetilde{T}^c \cap T_0} + h_{\widetilde{T}^c \cap T_0}\|_1 + \|x_{\widetilde{T}^c \cap T_0^c} + h_{\widetilde{T}^c \cap T_0^c}\|_1 
	  + \sum\limits_{j=1}^m\left(\omega_j \|x_{\widetilde{T}_j \cap T_0} + h_{\widetilde{T}_j \cap T_0}\|_1 + \omega_j\|x_{\widetilde{T}_j \cap T_0^c} + h_{\widetilde{T}_j \cap T_0^c}\|_1\right) \\ 
	& \quad \quad \quad \quad \quad \quad \quad \leq \quad \|x_{\widetilde{T}^c \cap T_0}\|_1 + \|x_{\widetilde{T}^c \cap T_0^c}\|_1 + \sum\limits_{j=1}^m\left(\omega_j\|x_{\widetilde{T}_j \cap T_0}\|_1 + \omega_j\|x_{\widetilde{T}_j \cap T_0^c}\|_1\right).
\end{array}
\end{equation*}
Next, we use the forward and reverse triangle inequalities to get
\begin{equation}\nonumber
	\sum\limits_{j=1}^m\left(\omega_j \|h_{\widetilde{T}_j \cap T_0^c}\|_1\right) + \|h_{\widetilde{T}^c \cap T_0^c}\|_1  \leq \|h_{\widetilde{T}^c \cap T_0}\|_1 + \sum\limits_{j=1}^m\omega_j \|h_{\widetilde{T}_j \cap T_0}\|_1  + 2\left(\|x_{\widetilde{T}^c \cap T_0^c}\|_1 + \sum\limits_{j=1}^m\omega_j \|x_{\widetilde{T}_j \cap T_0^c}\|_1\right).
\end{equation}
Adding $\sum\limits_{j=1}^m(1-\omega_j)\|h_{\widetilde{T}_j^c \cap T_0^c}\|_1$ on both sides of the inequality above we obtain
\begin{equation}\nonumber
\begin{array}{ll}
\sum\limits_{j=1}^m\|h_{\widetilde{T}_j \cap T_0^c}\|_1 + \|h_{\widetilde{T}^c \cap T_0^c}\|_1
	&\leq \sum\limits_{j=1}^m \omega_j \|h_{\widetilde{T}_j \cap T_0}\|_1 + \sum\limits_{j=1}^m(1-\omega_j) \|h_{\widetilde{T}_j \cap T_0^c}\|_1 +  \|h_{\widetilde{T}^c \cap T_0}\|_1\\
	 & \quad \quad + \quad 2\left(\|x_{\widetilde{T}^c \cap T_0^c}\|_1 + \sum\limits_{j=1}^m\omega_j \|x_{\widetilde{T}_j \cap T_0^c}\|_1\right).
\end{array}
\end{equation}
Since $\|h_{T_0^c}\|_1 =  \|h_{\widetilde{T} \cap T_0^c}\|_1 + \|h_{\widetilde{T}^c \cap T_0^c}\|_1$ and $\|h_{\widetilde{T} \cap T_0^c}\|_1 = \sum\limits_{j=1}^m \|h_{\widetilde{T}_j \cap T_0^c}\|_1$, this easily reduces to 
\begin{equation}\label{eq:weighted_optimality}
	\|h_{T_0^c}\|_1 \leq  \sum\limits_{j=1}^m \omega_j \|h_{\widetilde{T}_j \cap T_0}\|_1 + \sum\limits_{j=1}^m(1-\omega_j) \|h_{\widetilde{T}_j \cap T_0^c}\|_1 +  \|h_{\widetilde{T}^c \cap T_0}\|_1 + 2\left(\|x_{\widetilde{T}^c \cap T_0^c}\|_1 + \sum\limits_{j=1}^m\omega_j \|x_{\widetilde{T}_j \cap T_0^c}\|_1\right).
\end{equation}
Now consider the following term from the left hand side of \eqref{eq:weighted_optimality}
$$
\sum\limits_{j=1}^m \omega_j \|h_{\widetilde{T}_j \cap T_0}\|_1 + \sum\limits_{j=1}^m(1-\omega_j) \|h_{\widetilde{T}_j \cap T_0^c}\|_1 +  \|h_{\widetilde{T}^c \cap T_0}\|_1 
$$
Add and subtract $\sum_{j=1}^m (1-\omega_j)\|h_{\widetilde{T}_j^c\cap T_0}\|_1$, and since the set $\widetilde{T}_{j\alpha} = T_0 \cap \widetilde{T}_j$, we can write $\|h_{\widetilde{T}_j^c \cap T_0}\|_1 + \|h_{\widetilde{T}_j \cap T_0^c}\|_1 = \|h_{T_0 \cup \widetilde{T}\setminus \widetilde{T}_{j\alpha}}\|_1$ to get
$$
\begin{array}{ll}
&\sum\limits_{j=1}^m \omega_j \left(\|h_{\widetilde{T}_j \cap T_0}\|_1+\|h_{\widetilde{T}_j^c \cap T_0}\|_1\right)  + \sum\limits_{j=1}^m(1-\omega_j) \left(\|h_{\widetilde{T}_j \cap T_0^c}\|_1+ \|h_{\widetilde{T}_j^c \cap T_0}\|_1\right) +  \|h_{\widetilde{T}^c \cap T_0}\|_1 - \sum\limits_{j=1}^m \|h_{\widetilde{T}_j^c \cap T_0}\|_1 \\
= & \left(\sum\limits_{j=1}^m \omega_j\right)\|h_{T_0}\|_1+ \|h_{\widetilde{T}^c \cap T_0}\|_1 - \sum\limits_{j=1}^m \|h_{\widetilde{T}_j^c \cap T_0}\|_1  + \sum\limits_{j=1}^m(1-\omega_j) \|h_{T_0 \cup \widetilde{T}\setminus \widetilde{T}_{j\alpha}}\|_1 \\
= & \left(\sum\limits_{j=1}^m \omega_j - m + 1\right)\|h_{T_0}\|_1 + \sum\limits_{j=1}^m(1-\omega_j) \|h_{T_0 \cup \widetilde{T}\setminus \widetilde{T}_{j\alpha}}\|_1. 
\end{array}
$$
The last equality comes from $\|h_{T_0\cap \widetilde{T}_j^c}\|_1 = \|h_{\widetilde{T}^c \cap T_0}\|_1 + \|h_{T_0\cap\left(\widetilde{T}\setminus\widetilde{T}_j\right)}\|_1$ and $\sum\limits_{j=1}^m\|h_{T_0\cap\left(\widetilde{T}\setminus\widetilde{T}_j\right)}\|_1 = (m-1)\|h_{T_0\cap\widetilde{T}}\|_1$.

Consequently, we can reduce the bound on $\|h_{T_0^c}\|_1$ to the following expression:
\begin{equation}\label{eq:weighted_hT0c1}
	\|h_{T_0^c}\|_1 \leq  \left(\sum\limits_{j=1}^m \omega_j - m + 1\right)\|h_{T_0}\|_1 + \sum\limits_{j=1}^m(1-\omega_j) \|h_{T_0 \cup \widetilde{T}\setminus \widetilde{T}_{j\alpha}}\|_1 
	+ 2\left(\|x_{\widetilde{T}^c \cap T_0^c}\|_1 + \sum\limits_{j=1}^m\omega_j \|x_{\widetilde{T}_j \cap T_0^c}\|_1\right).
\end{equation}

Next we follow the technique of Cand{\`e}s et al.\cite{CRT05} and sort the coefficients of $h_{T_0^c}$ partitioning $T_0^c$ it into disjoint sets $T_j, j \in \{1,2,\ldots\}$ each of size $ak$, where $a > 1$. That is, $T_1$ indexes the $ak$ largest in magnitude coefficients of $h_{T_0^c}$, $T_2$ indexes the second $ak$ largest in magnitude coefficients of $h_{T_0^c}$, and so on. Note that this gives  $h_{T_0^c} = \sum_{j \geq 1} h_{T_j}$, with
\begin{equation}\label{eq:hTj2}
	\|h_{T_j}\|_2 \leq \sqrt{ak} \|h_{T_j}\|_{\infty} \leq (ak)^{-1/2} \|h_{T_{j-1}}\|_1.
\end{equation}
Let $T_{01} = T_0 \cup T_1$, then using (\ref{eq:hTj2}) and the triangle inequality we have
\begin{equation}\label{eq:hT01c2_sum_bound}
\begin{array}{lrl}
	&\|h_{T_{01}^c}\|_2 & \leq \sum\limits_{j \geq 2} \|h_{T_j}\|_2  \leq (ak)^{-1/2} \sum\limits_{j \geq 1} \|h_{T_j}\|_1\\
	&& \leq (ak)^{-1/2} \|h_{T_0^c}\|_1.
\end{array}
\end{equation}

Next, consider the feasibility of $x^\#$ and $x$. Both vectors are feasible, so we have $\|Ah\|_2 \leq 2\epsilon$ and
\begin{equation}\nonumber
\begin{array}{lrl}
 \|Ah_{T_{01}}\|_2 &\leq& 2\epsilon + \|Ah_{T_{01}^c}\|_2 
\leq 2\epsilon + \sum\limits_{j \geq 2} \|Ah_{T_j}\|_2 \\
& \leq& 2\epsilon + \sqrt{1+\delta_{ak}} \sum\limits_{j \geq 2} \|h_{T_j}\|_2.
\end{array}
\end{equation}
From \eqref{eq:weighted_hT0c1} and (\ref{eq:hT01c2_sum_bound}) we get
\begin{equation}\nonumber
\begin{array}{lrl}
& \|Ah_{T_{01}}\|_2 &\leq 2\epsilon + 2\frac{\sqrt{1+\delta_{ak}}}{\sqrt{ak}} \left(\|x_{\widetilde{T}^c \cap T_0^c}\|_1 + \sum\limits_{j=1}^m\omega_j \|x_{\widetilde{T}_j \cap T_0^c}\|_1\right)\\
&& \quad + \quad \frac{\sqrt{1+\delta_{ak}}}{\sqrt{ak}}\left((\sum\limits_{j=1}^m \omega_j - m + 1)\|h_{T_0}\|_1 + \sum\limits_{j=1}^m(1-\omega_j) \|h_{T_0 \cup \widetilde{T}\setminus \widetilde{T}_{j\alpha}}\|_1 \right).
\end{array}
\end{equation}
Noting that $|T_0 \cup \widetilde{T}\setminus \widetilde{T}_{j\alpha}| = (1 + \rho_j - 2\alpha_j\rho_j)k$,
\begin{equation}\nonumber
\begin{array}{lrl}
& \sqrt{1-\delta_{(a+1)k}}\|h_{T_{01}}\|_2 &\leq 2\epsilon + 2\frac{\sqrt{1+\delta_{ak}}}{\sqrt{ak}} \left(\|x_{\widetilde{T}^c \cap T_0^c}\|_1 + \sum\limits_{j=1}^m\omega_j \|x_{\widetilde{T}_j \cap T_0^c}\|_1\right)\\
&& \quad + \quad \frac{\sqrt{1+\delta_{ak}}}{\sqrt{a}}\left((\sum\limits_{j=1}^m \omega_j - m + 1)\|h_{T_0}\|_2 + \sum\limits_{j=1}^m(1-\omega_j)\sqrt{1 + \rho_j - 2\alpha_j\rho_j} \|h_{T_0 \cup \widetilde{T}\setminus \widetilde{T}_{j\alpha}}\|_2 \right).
\end{array}
\end{equation}
Since for every $j$ we have $\|h_{T_0
  \cup \widetilde{T}_j\setminus \widetilde{T}_{j\alpha}}\|_2 \leq
\|h_{T_{01}}\|_2$ and $\|h_{T_0}\|_2 \leq \|h_{T_{01}}\|_2$,
thus
\begin{equation}\label{eq:weighted_hT01_2}
	\|h_{T_{01}}\|_2 \leq \frac{2\epsilon + 2\frac{\sqrt{1+\delta_{ak}}}{\sqrt{ak}} \left(\|x_{\widetilde{T}^c \cap T_0^c}\|_1 + \sum\limits_{j=1}^m\omega_j \|x_{\widetilde{T}_j \cap T_0^c}\|_1\right)}{\sqrt{1-\delta_{(a+1)k}} - \frac{\sum\limits_{j=1}^m \omega_j - m + 1 + \sum\limits_{j=1}^m(1-\omega_j)\sqrt{1 + \rho_j - 2\alpha_j\rho_j}}{\sqrt{a}}\sqrt{1+\delta_{ak}}}.
\end{equation}

Finally, using $\|h\|_2 \leq \|h_{T_{01}}\|_2 + \|h_{T_{01}^c}\|_2$ and let $\gamma = \sum\limits_{j=1}^m \omega_j - m + 1 + \sum\limits_{j=1}^m(1-\omega_j)\sqrt{1 + \rho_j - 2\alpha_j\rho_j}$, we combine \eqref{eq:weighted_hT0c1}, (\ref{eq:hT01c2_sum_bound}) and (\ref{eq:weighted_hT01_2}) to get
\begin{equation}\label{eq:weighted_h2}
	\|h\|_2 \leq \frac{2\left(1+\frac{\gamma}{\sqrt{a}}\right)\epsilon + 2 \frac{
\sqrt{1-\delta_{(a+1)k}} + \sqrt{1+\delta_{ak}}}{\sqrt{ak}}\left(\|x_{\widetilde{T}^c \cap T_0^c}\|_1 + \sum\limits_{j=1}^m\omega_j \|x_{\widetilde{T}_j \cap T_0^c}\|_1\right)}{\sqrt{1-\delta_{(a+1)k}} - \frac{\gamma}{\sqrt{a}}\sqrt{1+\delta_{ak}}  },
\end{equation}
with the condition that the denominator is positive, equivalently 
$\delta_{ak} + \frac{a}{\gamma^2}\delta_{(a+1)k} < \frac{a}{\gamma^2} - 1$. 

\acknowledgments 
The authors would like to thank Rayan Saab for helpful discussions and
for sharing his code, which we used for conducting our audio
experiments. Both authors were supported in part by the Natural
Sciences and Engineering Research Council of Canada (NSERC)
Collaborative Research and Development Grant DNOISE II
(375142-08). {\"O}. Y{\i}lmaz was also supported in part by an NSERC
Discovery Grant.


\bibliography{sparse}   
\bibliographystyle{spiebib}   

\end{document}